\documentclass[lettersize,journal]{IEEEtran}
\usepackage{amsmath,amsfonts}
\usepackage{algorithmic}
\usepackage{algorithm}
\usepackage{array}
\usepackage[caption=false,font=normalsize,labelfont=sf,textfont=sf]{subfig}
\usepackage{textcomp}
\usepackage{stfloats}
\usepackage{url}
\usepackage{verbatim}
\usepackage{graphicx}
\usepackage{subcaption}
\usepackage{cite}
\usepackage{xcolor}
\usepackage{enumitem}

\begin{document}

\title{Saarthi: An End-to-End Intelligent Platform for Optimising Distributed Serverless Workloads}
\author{Siddharth~Agarwal,
        Maria~A.~Rodriguez,
        and~Rajkumar~Buyya%
\IEEEcompsocitemizethanks{\IEEEcompsocthanksitem The authors are with the Cloud Computing and Distributed Systems (CLOUDS) Laboratory, School of Computing and Information Systems, The University of Melbourne, VIC 3010, Australia.\protect\\
E-mail: {siddhartha@student.unimelb.edu.au,\\ \{maria.read,rbuyya\}@unimelb.edu.au}%
}}




\maketitle

\begin{abstract}
Function-as-a-Service offers significant advantages with its infrastructure abstraction, on-demand execution, and attractive no idle resource pricing for modern cloud applications. Despite these benefits, challenges such as startup latencies, static configurations, sub-optimal resource allocation and scheduling still exist due to coupled resource offering and workload-agnostic generic scheduling behaviour. These issues often lead to inconsistent function performance and unexpected operational costs for users and service providers. In particular, dynamic workload characteristics such as the number of inputs and their type and size are known to have a direct impact on a function's resource requirements, operational cost and execution time. This paper introduces \textit{Saarthi}, a novel, end-to-end serverless framework that intelligently manages the dynamic resource needs of function workloads, representing a significant step toward self-driving serverless platforms. Unlike platforms that rely on static resource configurations, Saarthi is input-aware, allowing it to intelligently \textit{anticipate resource requirements} based on the characteristics of an incoming request payload. This input-driven approach reinforces \textit{function right-sizing} and enables \textit{smart request orchestration} across available function configurations. Saarthi further integrates a \textit{proactive fault-tolerant redundancy mechanism} and employs a \textit{multi-objective Integer Linear Programming (ILP)} model to maintain an optimal function quantity. This optimisation aims to maximise system throughput while simultaneously reducing overall operational costs. We validate the effectiveness of Saarthi by implementing it as a framework atop OpenFaaS. Our results demonstrate Saarthi's ability to achieve up to 1.45x better throughput, 1.84x reduced costs, while maintaining up to 98.3\% service level targets with an overhead of up to 0.2 seconds as compared to the baseline OpenFaaS. 
\end{abstract}


\section{Introduction}
\label{sec:intro}

\IEEEPARstart{T}{he} Function-as-a-Service (FaaS) model of cloud computing promotes infrastructure abstraction, an on-demand execution and offers a compelling pay-per-use billing. FaaS lets developers focus on building application logic and deploy it as stateless containers or microVMs \cite{UCCSiddharth}, known as \textit{functions}. Functions often run for short durations and are executed in response to events, experiencing no idle costs with scale-to-zero capability. These advantages often make FaaS a popular choice for modern, scalable and cost-efficient cloud applications \cite{NetflixAWS}\cite{ServerlessVideo}. Several commercial and open-source FaaS platforms exist, including AWS Lambda \cite{awsLambda}, Azure Functions \cite{azureFunctions}, and OpenFaaS \cite{openfaas}, each with a distinct approach to providing serverless compute services. In FaaS, functions require compute resources for their execution, and developers are responsible for statically configuring them. Each platform offers a distinct set of configurable resources and runtime settings that are applied to the functions at the time of deployment. These configurations include resources such as CPU, memory, and temporary storage, as well as tunable parameters like the maximum execution time of a function. Generally, platforms such as AWS Lambda allocate CPU and other compute resources proportional to the configured memory. When a function is invoked, platforms utilise this configuration to scale function instances seamlessly. 

It is a well-researched fact that the amount of configured resources significantly impacts the performance of functions \cite{UCCSiddharth}\cite{ParrotFish}, affecting both execution time and operational cost. Consequently, determining function resources emerges as a pivotal deployment decision that directly influences the user's perceived latency and costs incurred by developers. For example, allocating more resources than required improves function execution time, but it may lead to increased operational cost. Conversely, under-provisioning a function's resources can throttle its performance, potentially leading to failed executions. Furthermore, a function's resource requirements are not static, as they dynamically vary with the workload characteristics such as input payload, its data type, size, and complexity \cite{OFC}. This makes a single static configuration inefficient across different executions, as a setting optimised for one workload may be significantly insufficient or over-provisioned for another. Since FaaS platforms often do not provide service-level guarantees \cite{SLAM} for execution time, developers resort to extensive manual profiling to determine optimal configurations. This introduces a significant overhead in terms of both developer time and operational cost, as the process is tedious, error-prone, and rarely accounts for the full range of dynamic inputs. 

However, acknowledging input-aware 
memory variability leads to a paradigm where multiple function versions with different configurations co-exist. This input and version-aware function management makes scheduling decisions non-trivial and introduces a new set of operational challenges. It is impractical to instantiate a new function instance for every incoming input 
as it incurs an expensive cold start overhead. Additionally, frequent creation of distinct function versions leads to excessive resource fragmentation for the service provider and impacts their ability to efficiently make scheduling and placement decisions \cite{WithGreatFreedomComesGreatOpportunity}. Therefore, it only makes sense to strike a balance between reusing existing versions with sufficient resources or deploying new function versions tailored to the incoming workload. This strategy ensures that the system not only adapts to the dynamic workload needs, but also reduces redundant overheads with minimal developer intervention while keeping the user latency and provider costs in check.



To address the aforementioned function configuration challenge, a number of recent works \cite{UCCSiddharth}\cite{ParrotFish}\cite{StepConf}\cite{Cypress} offer solutions that can be utilised to optimise operational cost or performance guarantees for different workloads. While \cite{StepConf} provides a cost-optimised function parallelism for workflows, \cite{ParrotFish} determines optimal working memory range in a cost-efficient way. On the other hand, \cite{UCCSiddharth} exploits the workload characteristics to furnish input-aware function configurations that reduce operational cost and allocated resources. Although existing studies provide independent approaches to recommend or determine the optimal resources based on distinct objectives, they either fail to orchestrate the workload to the desired configuration 
or do not weigh the benefits of reusing existing versions. Additionally, scholarly works \cite{OFC}\cite{Llama}\cite{charmseeker} explore application-specific configurable parameters, such as video frames or model hyperparameters, to propose an optimal value of these settings. However, there exists no single, automated framework that can manage dynamically changing function requirements and simultaneously provide optimisation for both performance and cost while ensuring fault-tolerance. Additionally, the manual effort involved in profiling and setting the resource configuration warrants an adaptive approach to ensure consistent performance and cost efficiency. This research gap highlights the need for an integrated, end-to-end solution that can intelligently orchestrate function requests and resources to meet complex service-level requirements, leading towards \textit{self-driving} serverless platforms. The idea of self-driving platforms is to completely abstract the resource management tasks from the function lifecycle and enable the developers to truly focus on building business logic. 


This paper proposes \textbf{Saarthi}, an end-to-end FaaS framework that leverages input-aware resource predictions to adaptively orchestrate incoming workloads, optimise the number of active functions and ensure service continuity by compensating for potential function failures. It introduces a prediction module to determine the input-aware resource requirements of an incoming request. An adaptive request balancer that intelligently re-uses an existing version or cold-starts a new one. Alongside, an independent ILP-based function optimiser is proposed that continuously monitors and maintains a healthy and desired cluster state. Furthermore, a redundancy-based scaling component provides fault-tolerance by compensating for function failures. We build Saarthi atop open-source OpenFaaS \cite{openfaas} serverless framework and offer different modules as standalone services acting in unison. We experiment with a set of functions from standard benchmark suites \cite{FuncitonBench} \cite{SebsBenchmark} to evaluate their performance with the Saarthi framework. We further evaluate the impact on operational cost of the functions and compare it with the baseline OpenFaaS platform. Additionally, an ablation study is done with different components of the Saarthi framework to identify the overheads introduced by individual services and highlight their overall benefit. In doing so, we make the following contributions:
\begin{itemize}
    \item We propose a hybrid optimisation for request orchestration that combines a heuristic scoring with probabilistic element. It balances between exploration and exploitation to either use immediately available function versions or occasionally cold starting new instances. 
    \item We propose a fault-tolerant request handling with a $G/G/c/K$ buffer to temporarily queue requests and prevent immediate dropping of requests with a retry mechanism to improve system resilience. 
    \item We propose an ILP-based function optimiser that continuously monitors and adjusts the number of function instances, while adapting to dynamic workloads and operational costs.
    \item We employ a redundancy-based, fault tolerant function scaling mechanism to ensure service continuity and faster service recovery by mitigating mis-predictions and instance failures.
    \item We incorporate the discussed components into \textit{Saarthi}, a complete framework to automate and integrate multiple aspects of serverless function management, and evaluate its performance against the baseline OpenFaaS using the established benchmark functions \cite{SebsBenchmark}\cite{FuncitonBench}.
\end{itemize}

\section{Motivation}
\label{sec:motivation}



\begin{figure}[th]
    \centering
    \includegraphics[height=4.2cm]{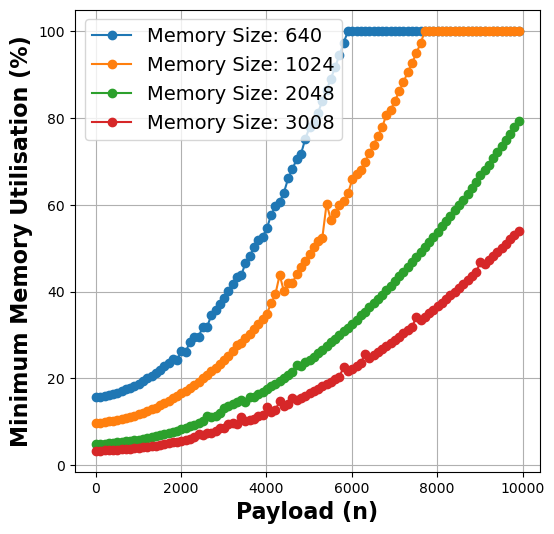}%
    \includegraphics[height=4.2cm]{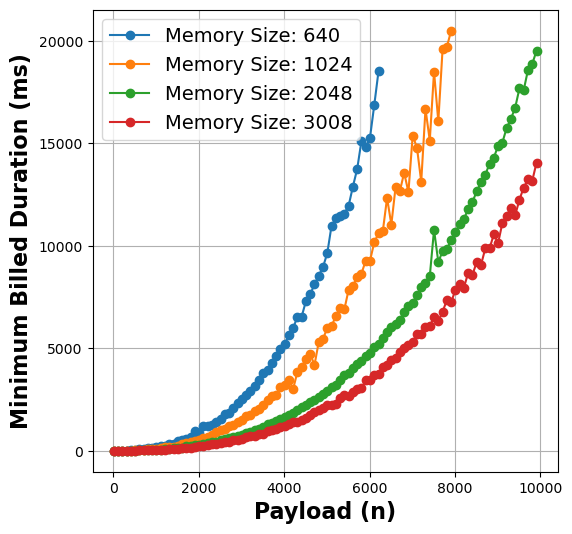}
    \caption{Comparison of Input Payload vs Memory Utilisation (left) and Billed Execution Duration (right).}
    \label{fig:memory-payload}
\end{figure}

The motivation for building Saarthi as a complete function lifecycle manager stems from the inherent shortcomings of the current FaaS platforms, which generally fail to deliver the operational simplicity and efficiency of serverless. Although FaaS introduces certain benefits, these platforms still exhibit critical limitations that lead to inconsistent performance, high operational cost and significant developer overhead. Additionally, these inefficiencies result in resource fragmentation, sub-optimal resource allocation and scheduling, and high maintenance costs for providers. One of the shortcomings is the workload-agnostic nature of most FaaS platforms \cite{scSOTAli}. They treat every function invocation in the same way, regardless of the distinct workload characteristics such as input parameters and set a static, one-size-fits-all resource configuration.
Existing studies \cite{UCCSiddharth}\cite{ParrotFish}\cite{OFC} discuss the effect of workload characteristics, in addition to allocating proportional CPU, network bandwidth and other compute resources to memory on function performance and highlight the complex relation between input payload and overall performance. Therefore, a configuration optimised for one type or size of input will either over-provision the resources for a simple request, leading to economic waste, or under-provision for a complex request causing a degraded performance and execution failure in the worst case. Fig. \ref{fig:memory-payload} demonstrates the performance of a compute-intensive function, \textit{linpack}, from \cite{FuncitonBench} and plots its memory utilisation and execution time for various input payloads. Fig. \ref{fig:memory-payload} (left) shows that memory utilisation of a function varies with input payload values at different memory settings and consequently affects the execution time (right). Thus, a memory configuration of 640 MB that ensures a successful execution of input payload, $n$ (number of linear equations to solve), up to 6000, results in failed execution for larger inputs, Fig. \ref{fig:memory-payload} (right). On the other hand, setting the memory configuration to 3008 MB leads to highly under-utilised resources for smaller inputs, however, it improves the execution duration of the function.

Another limitation of current FaaS platforms is the lack of service-level guarantees for performance metrics such as latency or throughput \cite{StepConf}\cite{SLAM}. Although providers such as AWS and Azure typically cover infrastructure availability and reliability \cite{AWSSLA}, they do not explicitly ensure function service-level objectives (SLOs). The lack of predictable performance makes it difficult for developers to build and deploy latency-sensitive applications. In Fig. \ref{fig:memory-payload} (right), if we consider the execution time SLO as $5$ seconds, then for some invocations i.e., smaller inputs ($n \leq 4000$), 640 MB memory setting would suffice, however, larger memory configurations such as 2048 MB or 3008 MB would be required for other invocations to successfully execute within the threshold. In addition to this, current platforms do not guarantee that functions run within a specified cost budget \cite{COSE.V1}\cite{COSE.V2}, while meeting execution time thresholds. FaaS billing 
creates a critical trade-off where a function with a larger, more expensive memory setting might run for a shorter time, potentially resulting in a lower overall cost than a function with a smaller, cheaper setting that executes for longer. This forces developers to take on the heavy burden of manual profiling to find the 'right' resource configuration, a time-consuming and often inaccurate process. Consequently, this overhead defeats the purpose of the serverless paradigm, which is to abstract away such execution complexities and avoid developer overheads. This highlights the need for an intelligent framework that can navigate these complex, often conflicting objectives. 

The challenge of resource configuration in FaaS is equally pressing for cloud service providers (CSP). When users manually specify resource allocations, often by overestimating to avoid timeouts, or underestimating to save on perceived costs, it leads to suboptimal utilisation across the provider's shared infrastructure. Such mis-configurations cause \textit{resource fragmentation}, making efficient bin-packing and function scheduling difficult at scale \cite{berkleyView2019}\cite{Sizeless}. This wastage of physical resources reduces overall system efficiency, drives up operational costs, and complicates performance prediction and service quality assurance. In particular, the hard-to-predict resource requirements lead to design trade-offs for startup speeds, hardware isolation and cost efficiency in the cloud \cite{Firecracker}. Providers are thus motivated to move beyond the developer-driven model towards 
automated resource management, which optimises infrastructure use while safeguarding SLOs for all tenants. Following the idea of tailored resource configuration for incoming function input, the frequent start-up of distinct function versions adds further complexity to the fragmentation challenge. This is because these versions, coupled with the dynamic idle timeout for each function, can lead to numerous simultaneously active or idle instances. Given this, it makes little sense to always create a new instance based on every workload's resource requirement. Instead, it is warranted to first check for any existing, idle version that can potentially fulfil the SLO, and use that instance to avoid a cold start and save provider costs while maintaining the defined SLOs. This highlights the need for a trade-off approach that can consider this overhead and make appropriate, informed decisions. 


Furthermore, a service provider must also consider the function concurrency limits and instance count for individual versions. Typically, for platforms like OpenFaaS, function concurrency or the number of concurrent requests a single instance can serve, plays a critical role in resource consumption and utilisation \cite{OpenFaaSAutoScaling}. The rationale here is that instead of creating many small instances to serve diverse workloads, it may be more efficient to consolidate these onto a smaller number of larger, more capable instances that can serve multiple workload types within the same SLO limits. This gives the provider greater control over resource fragmentation, utilisation, and allocation. Therefore, to fulfil both cost and performance goals, the platform needs a multi-objective optimisation approach that can proactively understand the workload and maintain the right number of function instances that complement the platform's ability to auto-scale.

These limitations highlight a critical research gap: the absence of a single, end-to-end framework that can holistically and intelligently manage the entire function lifecycle. An effective solution must be able to autonomously adapt to dynamically changing function requirements, balance the competing objectives of performance and cost, and ensure fault-tolerance without manual intervention. Saarthi is designed to fill this gap, moving beyond basic function execution to provide a comprehensive, self-driving serverless solution.


\section{SAARTHI System Architecture}
\label{sec:saathi}

\begin{figure}
    \centering
    \includegraphics[width=1\linewidth]{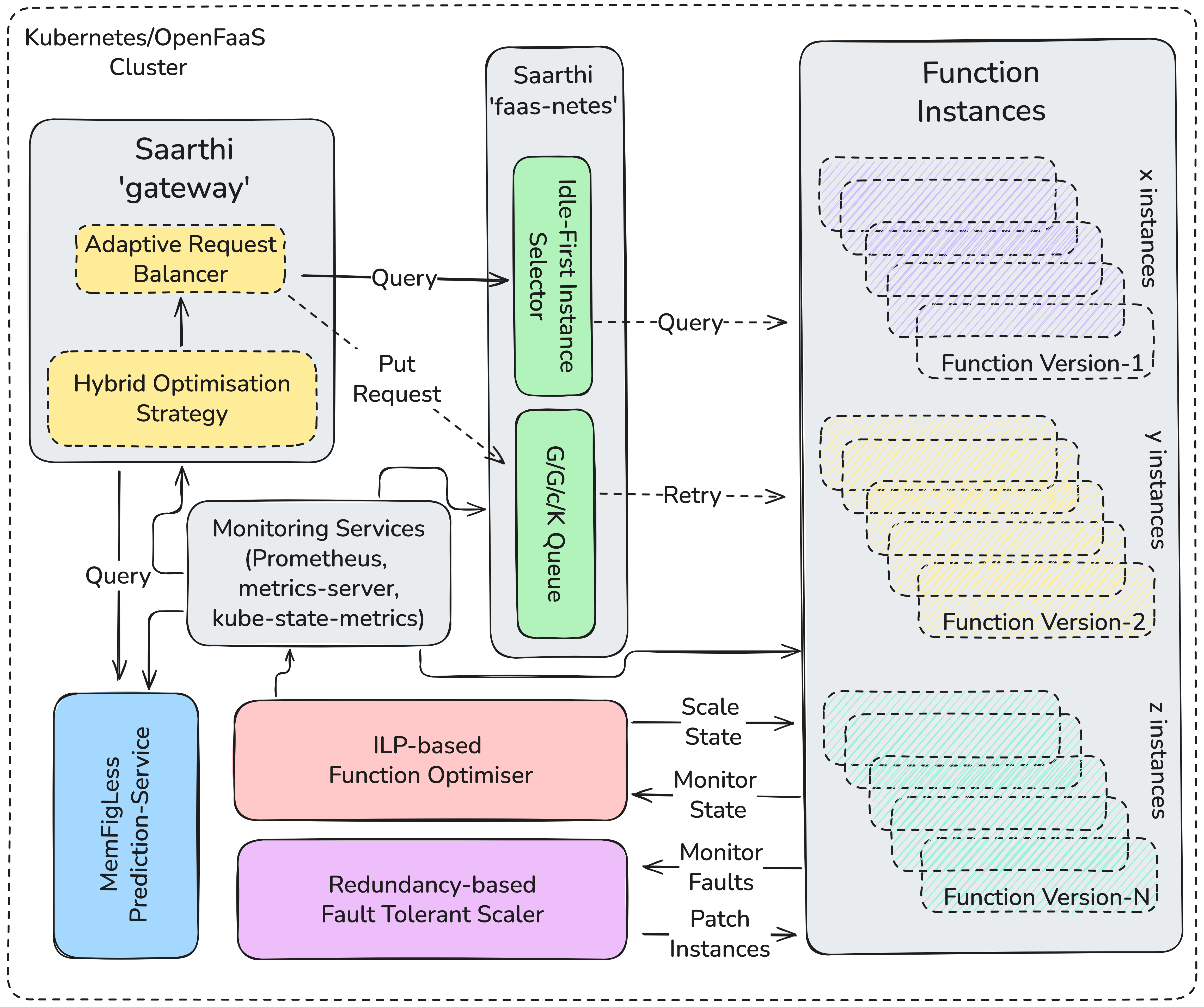}
    \caption{Saarthi System Architecture}
    \label{fig:architecture}
\end{figure}

The core of Saarthi is built on top of Kubernetes-based OpenFaaS \cite{openfaas} framework, whose two critical components Gateway and Kubernetes provider (\textit{faas-netes}) ensure the serverless experience. Saarthi introduces a number of loosely coupled independent components, Fig. \ref{fig:architecture}, to provide a self-driving serverless platform where input-aware resource configuration decisions are automatically handled for the workload. 

\subsection{Saarthi Request Lifecycle: A Conceptual Overview}
To understand how Saarthi works, it's helpful to follow the journey of a single function request. When a request arrives, the system first utilises a Prediction Service to determine the function's resource needs based on the request input/payload. This information is then passed to an Adaptive Request Balancer, which acts as the system's primary decision-maker. The balancer makes a crucial choice: should it reuse an existing, \textit{warm} function instance or should it create a new one to handle the request? This decision is informed by a hybrid optimisation strategy. 
If the balancer is unable to immediately find a suitable instance, it places the request in a queue to prevent immediate failure. While the balancer handles individual requests, a separate ILP-based Optimisation Engine continuously monitors the entire cluster to manage the overall pool of function instances to meet long-term performance and cost objectives. Finally, a proactive Fault-Tolerance Mechanism acts as a safety guardrail, detecting and compensating for failures to ensure the service remains available and reliable. By orchestrating the flow of requests and resources through these interconnected components, Saarthi automates the entire function lifecycle.

\subsection{Input-Aware Prediction Service}
\label{subsec:memfigless}

Saarthi's goal of creating a self-driving serverless platform builds on a deep understanding of workload characteristics. As established by prior research \cite{Cypress}\cite{WithGreatFreedomComesGreatOpportunity}, a function's performance is critically tied to its resource configuration, which in turn is heavily influenced by the request input/payload. To enable automated decisions for both request orchestration and resource allocation, Saarthi incorporates a Prediction Service.

This service takes advantage of input-aware ensemble learning models, an online Random Forest Regression (RFR) pipeline, adapted from previous work \cite{UCCSiddharth}. This allows Saarthi to perform real-time input-aware inference by predicting an incoming request's ($Q$) resource needs, $R_p$ before it is executed. The details of RFR and performance are discussed in prior work \cite{UCCSiddharth}. Furthermore, our training workflow supports incremental learning, which means the Prediction Service can be configured with a \textit{refresh interval} to continuously synchronise with and utilise the latest, most accurate models. This ensures that Saarthi remains adaptive and efficient as workload patterns evolve over time.  

\subsection{Adaptive Request Balancer and $G/G/c/K$ Queue}
\label{subsec:version-fallback}

Once the platform determines the input-aware resource requirement, $R_p$, of the incoming request $Q$, an \textit{Adaptive Request Balancer} (ARB) comes into play that is integrated into OpenFaaS \textit{Gateway}. Saarthi leverages this component to orchestrate the request to an appropriate function version. This orchestration mechanism is a critical feature, as it highlights a fundamental distinction between OpenFaaS and commercial FaaS platforms such as AWS Lambda. While these platforms achieve fine-grained, per-request scalability through automated, infrastructure-level provisioning for nearly every incoming event, OpenFaaS remains inherently constrained by the scaling semantics of the underlying Kubernetes architecture. OpenFaaS cannot technically scale for every individual request due to a control-plane overhead, database lock contention in \textit{etcd} 
and scheduler back-pressure stemming from phenomena like the \textit{thundering herd problem} when many scaling events occur close together. Even in commercial OpenFaaS Pro plans that offer advanced scaling metrics such as “capacity mode” \cite{OpenFaaSAutoScaling}, these advanced approaches still operate atop standard Kubernetes primitives and remain subject to underlying resource and scheduling bottlenecks. As a result, OpenFaaS often exhibit inefficient resource utilisation and can demonstrate suboptimal performance. For example, it could lead to delayed autoscale reactions or an inability to immediately allocate capacity for spikes of concurrent requests \cite{OpenFaaSQueue} as compared to managed solutions like AWS Lambda.


To overcome these limitations, Saarthi's ARB employs a \textbf{hybrid optimisation strategy} that balances exploration and exploitation of the current function instances, $F$, based on the acquired knowledge of resource requirements, $R_p$. First, a heuristic scoring algorithm identifies the most efficient function version to handle the request input, typically preferring a version with predicted configuration, $f_{exact}$ or the smallest version with sufficient resources among the other available versions, $F_{available}$. This is achieved by calculating a score for each function version based on the difference between its resources and the predicted requirements. This is a critical feature as Saarthi's input-aware design often deploys multiple function versions to handle different workload characteristics. Therefore, this process efficiently \textit{exploits} already existing ("warm") function versions by prioritising their reuse over creating new ones, leading to reduced cold-starts. However, to avoid a long-term sub-optimal state, the system occasionally introduces a probabilistic decision-making step with a pre-defined tolerance (e.g., 20\%) to \textit{explore} cold starting an instance with the predicted resources. This strategy improves upon both the standard OpenFaaS community/professional autoscaling \cite{OpenFaaSComparison}, which often relies solely on trailing metrics like requests-per-second (RPS) or simple in-flight request counts, and naive greedy schemes which can rapidly overload available functions, leading to queuing, timeouts, or overloaded instances. In contrast, the probabilistic exploration path proactively cold-starts new instances even in the absence of immediate pressure, ensuring balanced distribution and adaptiveness for real workload surges. By doing so, Saarthi mitigates the risk of local minima where requests are routed only to a saturated set of function instances, a frequent challenge faced by both baseline OpenFaaS scaling and conventional heuristics. An overview of request orchestration is demonstrated in Algorithm \ref{alg:saarthi_orchestration}.

\begin{algorithm}[H]
\caption{Saarthi Adaptive Request Balancer}
\label{alg:saarthi_orchestration}
\begin{algorithmic}[1]
\REQUIRE Incoming request $Q$ for function $S_{name}$ , predicted configuration $R_p$ 
\ENSURE Reuse or start a function version

\STATE Define $F$ as the set of all existing function instances.

\STATE $f_{exact}$ $\leftarrow$ an idle instance, $p_i^{R_p}$ of exact configuration
\IF{$f_{exact}$ exists}
\STATE Route request to function version $S_{name}^{R_p}$ 
\RETURN
\ENDIF

\STATE $F_{available}$ $\leftarrow$find all available alternative function versions
\IF{$F_{available}$ is not empty}
\STATE $f_{best}$ $\leftarrow$select the instance from $F_{available}$  with the lowest S($f_{best}$ )
\STATE $S_{best}$ $\leftarrow$S($f_{best}$ )
\STATE $S_{CS}$ $\leftarrow$random score from a ±20\% window of $S_{best}$ 
\IF{$S_{CS}$ $\leq$ $S_{best}$ }
\STATE \COMMENT{Explore: Cold start is better}
\STATE Deploy new function version $S_{name}^{R_p}$ 
\STATE Route request to new version $S_{name}^{R_p}$
\RETURN
\ELSE
\STATE \COMMENT{Exploit: Use the existing best version}
\STATE Route request to function version of $f_{best}$ 
\RETURN
\ENDIF
\ENDIF

\end{algorithmic}
\end{algorithm}

Consequently, the balancer's high-level decision to forward a request to a function version, $S_{name}^{R_p}$, is based on the availability of its idle instances, $p_i^{R_p} \in \{1 \dots P^{R_p}\} $. The rationale for balancing between versions arise from the fact that the platform initially determines the input-aware resource configuration, leading to simultaneously active distinct function versions. This approach reduces the decision overhead at the Gateway, allowing it to remain decoupled from the Kubernetes provider-level logic and allowing it to manage concurrent requests more effectively. The final selection of a specific function instance is delegated to the \textit{faas-netes} component which employs an \textbf{idle-first instance} selection approach to serve the request. An instance, $p_i^*$ is considered idle if its concurrently active requests, $C_p$ is less than its configured function concurrency, $M_p$. The idle-first heuristic operates on two-stages where it first identifies all the idle instances for a function version and then attempts to claim one of them. In doing so, the system atomically adjusts the active connections counter of the selected instance, $p_i^*$ via \textit{optimistic locking} mechanism where the request is only routed upon successful claim. The system re-tries with an updated set of idle instances if the claim fails due to a race condition and queues it if no idle instances can be claimed after a pre-defined number of re-tries. 

When multiple requests simultaneously reach the same function version or if the system is unable to find an idle instance, a $\textbf{G/G/c/K}$ \textbf{queue} is used to avoid immediate failures. The queue represents a generalised system where the request arrival rate, $G$ and processing rate, $G$ is unpredictable while accounting for a finite number of function instances, $c$ and a limited buffer size, $K$. The queuing of requests also align with the best-effort execution model, as some instances of a respective function version may become idle by the time a request is re-tried from the queue. This adaptive approach not only maximises resource utilisation and keeps latency low within a standard Kubernetes setup, but it also adapts to changing workloads. In doing so, it attempts to provide a robust request handling without being restricted by the typical limitations of Kubernetes-based FaaS.

\subsection{ILP-Based Optimisation Engine}
\label{subsec:optimiser}

While the ARB targets the immediate handling of incoming requests, the system must manage the overall operational capacity to meet its long-term performance and cost objectives. To address this, Saarthi introduces an \textit{ILP-based Optimisation Engine} to maintain the required number of function instances. A multi-objective ILP optimisation \cite{WithGreatFreedomComesGreatOpportunity} emerges as a natural choice for this task for several key reasons. First, it provides a robust mathematical model to optimise contrasting objectives, such as minimising cost while maximising performance \cite{OptimizingMemoryAllocationinaServerless}. Second, the decision variables in the model are inherently integers \cite{CostMinimisationSC}, as they represent the number of function instances, $x_{f_v}$ and the number of requests to be assigned, $y^r_{f_v}$ (used for internal calculation). Lastly, and crucially, the objective functions and all constraints, including operational cost, execution time, and throughput, can be accurately modelled as a linear function of these integer decision variables \cite{COSE.V2}.

The optimisation engine runs as a standalone component, executing at a configurable interval (e.g., every minute) to ensure the long-term health of the function cluster. It complements the per-request decisions of ARB by taking a comprehensive, cluster-wide view of resources. By default, it follows a greedy approach, monitoring the workload from the last interval and making decisions based on this historical knowledge. This ensures a fast and efficient response to immediate changes. We formulate the objective of maintaining an optimal number of function instances, maximising throughput while minimising operational cost, as a multi-objective global optimisation problem, as follows:

\begin{multline}
\min \Big[
\sum_{f_v} \alpha \cdot x_{fv} \cdot cost_{fv}
+ \sum_{r} \beta \cdot (demand_r - served_r) \cdot penalty_r \\
- \sum_{r} \gamma \cdot served_r \cdot utility_r
\Big]
\end{multline}

;where $\alpha, \beta,\gamma$ are cost and utility parameters. To achieve its objectives, the component runs an ILP optimisation loop that calculates a combination of three distinct elements. 
First is the operational cost of all deployed instances ($\sum_{f_v} x_{f_v} \cdot cost_{f_v}$) of function version $f_v \in \{f_1 \dots f_n\}$ with execution cost $cost_{f_v}$. Second is a penalty for an unserved demand, i.e., $\sum_r (demand_r - served_r) \cdot penalty_r$ that models the cost of service failure, such as a missed service level objective thresholds for requests $r \in demand_r$ of a specific configuration. Finally, it accounts for the value (or utility) gained from every served request ($\sum_r served_r \cdot utility_r$), which motivates the optimiser to prioritise high-utility tasks. To arrive at a solution, the optimiser uses decision variables to determine the optimal number of instances, $x^{*}_{f_v}$ to deploy or maintain for each function version and how to assign incoming requests $y^{r}_{f_v} \leq demand_r$ to them. These decisions are governed by a set of constraints that ensure the total CPU ($\sum x_{f_v} \cdot cpu_{f_v} \leq C_{cpu}$) and memory ($\sum x_{f_v} \cdot mem_{f_v} \leq C_{mem}$) used do not exceed the cluster's capacity. Additionally, it ensures that the number of requests assigned to any instance does not exceed its processing capacity $y_{f_v}^r \leq M_{f_v}$, i.e., the function concurrency. The optimiser's intelligence is particularly evident in its ability to handle trade-off scenarios, such as determining when a few larger function versions are more beneficial than maintaining numerous smaller ones. This strategy not only improves resource efficiency but also reduces resource fragmentation across the cluster. 
Furthermore, the optimiser is constrained by end-user limits as it conforms to a best-effort execution model when these constraints are exceeded. This ensures that the system performs the best possible optimisation within its given limits, prioritising overall system stability.

\begin{algorithm}[H]
\caption{Fault-Tolerant Redundancy Algorithm}
\label{alg:redundancy}
\begin{algorithmic}[1]
\REQUIRE Set of functions $F$, Cooldown period $T_{cooldown}$
\ENSURE Scaling actions for each function

\FORALL{function $f_i$ $\in$ $F$}
\IF{Time since last scale action for $f_i < T_{cooldown}$ }
\STATE Continue to next function
\ENDIF
\STATE failingPods $\leftarrow$ Count pods of $f_i$  with status $S_{fail} \in \{OOMKilled, CrashLoopBackOff\}$
\IF{failingPods $>$ 0}
\STATE currentReplicas $\leftarrow$ Get current replicas for $f_i$ 
\STATE desiredReplicas $\leftarrow$ currentReplicas $+$ failingPods
\STATE Scale $f_i$  to desiredReplicas
\STATE Record current time as last scale action for $f_i$ 
\ENDIF
\ENDFOR
\end{algorithmic}
\end{algorithm}

\subsection{Fault-Tolerant Redundancy Mechanism}
\label{subsec:error-scaling}

As discussed in Sec. \ref{subsec:version-fallback}, a function instance can be configured with concurrency, $M_p$, meaning it can serve multiple requests simultaneously by sharing its resources. The system primarily determines the function version based on predicted resource requirements from the adapted models, Sec. \ref{subsec:memfigless}, to route the incoming request, $Q$. However, when a mis-prediction happens and a request is directed to an under-provisioned function version, it can lead to a function failure. This has a cascading effect where other active requests are also terminated due to the instance's unavailability, overloading other available function instances with incoming workload. 



To prevent this, Saarthi employs a proactive, fault-tolerant redundancy mechanism, Algorithm \ref{alg:redundancy}, that continuously monitors at a configured \textit{interval} for early signs of service degradation $S_{fail}$, such as high memory usage. When a potential failure is detected, this mechanism autonomously compensates for the failing instances by adding additional function units, i.e., a simple \textit{additive scaling strategy}. This rapid, proactive scaling helps to absorb the sudden surge of traffic and prevent a thundering herd problem where a large number of requests overwhelm the remaining, already burdened instances. To ensure stability and prevent unnecessary thrashing of resources, the component also implements a cool-down period, $T_{cooldown}$, which serves as a safeguard against conflicting scaling operations from the optimisation engine.

\section{Performance Evaluation}
\label{sec:evaluation}

In this section, we present Saarthi's implementation, along with the experimental setup and component configuration parameters. We discuss our assumptions and perform an experimental study to quantify the contributions of Saarthi's components to function performance. Our findings demonstrate that by working in coherence these components significantly impact key metrics such as operational cost, execution latency, and resource utilisation when compared to baseline OpenFaaS-Community Edition (CE).

\subsection{Implementation and System Setup}
\label{subsec:setup}
We implement the core components of Saarthi as an extension to the OpenFaaS-CE \cite{openfaas}, leveraging a multi-language microservices architecture to ensure scalability and modularity. It utilises the underlying Kubernetes (v1.32.8) cluster with an overall capacity of 68 vCPUs and 288 GB of memory, spread across 6 virtual nodes. For research purpose only, we override the OpenFaaS defaults to enable maximum 50 distinct function deployments scaling up to 100 instances, each and configuring longer timeout of 10 minutes for a function execution. Saarthi's primary components -  adaptive request balancer with a hybrid optimisation strategy, coordinating with an idle-first instance selection approach, and a waiting queue, are implemented in \textit{Go}. They are seamlessly integrated into the OpenFaaS \textit{gateway} and the \textit{faas-netes} provider (both configured with 100 millicore vCPU and 200 MiB memory), utilising programming language primitives for high-performance concurrency and thread management. The $G/G/c/k$ queue is highly configurable, allowing for fine-grained control over the frequency and duration of its retry attempts. The redundancy-based fault tolerance logic operates as a lightweight \textit{subroutine} alongside the main \textit{gateway} process to proactively monitor failures at regular configurable intervals. The prediction models, adapted from the previous work \cite{UCCSiddharth}, are exposed as dedicated API-based \textit{Python} services configured with sufficient resources to provide real-time inference at scale. The optimisation workflow is implemented as two coordinating services, a monitoring engine and an ILP optimiser. The monitoring engine (300 millicore vCPU and 256 MiB memory) is responsible for interacting with services such as \textit{metrics-server}, \textit{kube-state-metrics} and \textit{Prometheus} to collect the current state of the cluster and
a \textit{Python}-based ILP optimiser \footnote{The ILP optimisation is implemented using PuLP \cite{pulp} library} (250 millicore vCPU and 512 MiB memory) that consumes the monitored cluster state and runs a complex ILP optimisation loop to determine the desired state.

We evaluate the performance of Saarthi on a variety of serverless functions from various benchmarking suites \cite{FuncitonBench}\cite{SebsBenchmark}, including CPU/memory intensive (\textit{matmul, linpack, pyaes}), scientific functions (\textit{graph-mst, graph-bfs}) and dynamic website generation (\textit{chameleon}) by conducting an ablation study. The functions are developed in Python v3.12 and deployed using the official \textit{python3-http} template from OpenFaaS. The template utilises OpenFaaS \textit{watchdog} v0.10.7, a critical component to enable concurrent handling of requests. The resource configuration and function payload range is adapted from previous work \cite{UCCSiddharth} and we utilise the AWS Lambda pricing \cite{awsLambdaPricing} for cost calculations. To compare and validate the performance of proposed components, we analyse the open-source Azure function traces \cite{AzureTraces} and select unique workload patterns associated with different function events corresponding to every deployed function. Additionally, conforming to the \textit{log-normal} distribution of workloads \cite{AzureTraces}, we generate a \textit{log-normally} distributed function input values and utilise \textit{Poisson} distribution for request inter-arrival times to closely match the real-time traffic and application payloads.

All Saarthi components are designed to be highly configurable, allowing for fine-grained control over their behaviour. For instance, the prediction service can be configured with a \textit{refresh interval} (default 2 hours) to actively synchronise with the latest learned models. The optimisation engine exposes a decision-making frequency (default every 1 minute) and allows future extension endpoints for optional services such as workload prediction or pricing. Additionally, it explicitly allows configuration of cold-start trade-offs when optimising function deployments to account for prediction-based workload optimisation, disabled by default due to greedy, history-based workload optimisation. To prevent overwhelming a function instance, we set the function concurrency to 10 and similarly, we fix the $G/G/c/K$ queue length to 10 with a retry interval of 10 milliseconds. This modular design of components further allow us to fine-tune the platform for application specific optimisation, if required. 

\subsection{Experiments }
\label{subsec:ablation}

To evaluate the performance of our proposed solution, we conducted a comparative performance study with different versions of the Saarthi framework. The objective of this study is to measure the impact of key components on overall function and platform performance. For this study, all results are measured and evaluated against the baseline OpenFaaS-CE with its default settings and a resource configuration of 1769 MiB memory and 1 vCPU, consistent with standard AWS Lambda \cite{awsLambdaQuotas} quotas. We use a consistent set of functions, workloads, and metrics, including execution latency, operational cost, and overall system resource usage for comparison. No tests were performed in isolation to mimic a multi-tenant environment, which demonstrates the practical benefits of our framework's components when they work together in a realistic setting. The prior work \cite{UCCSiddharth} discusses the value of input-based memory configuration, and therefore, we do not re-iterate those results.

To strengthen Saarthi's position, we compare its performance against two different input-aware versions: Saarthi-MVQ, a version with the core request orchestration and a generalised waiting queue, and Saarthi-MEVQ, which adds the fault-tolerance mechanism. The full Saarthi framework with all components enabled is referred to as Saarthi-MOEVQ in our experiments. We selected different function triggers, such as \textit{http} and \textit{orchestration}, from available traces \cite{AzureTraces} to cover a variety of real-world usage patterns. Consequently, we conducted the experiments for 120 minutes of workload data corresponding to diverse functions and present our findings.

\begin{figure}[h]
    \centering
    \includegraphics[width=1\linewidth]{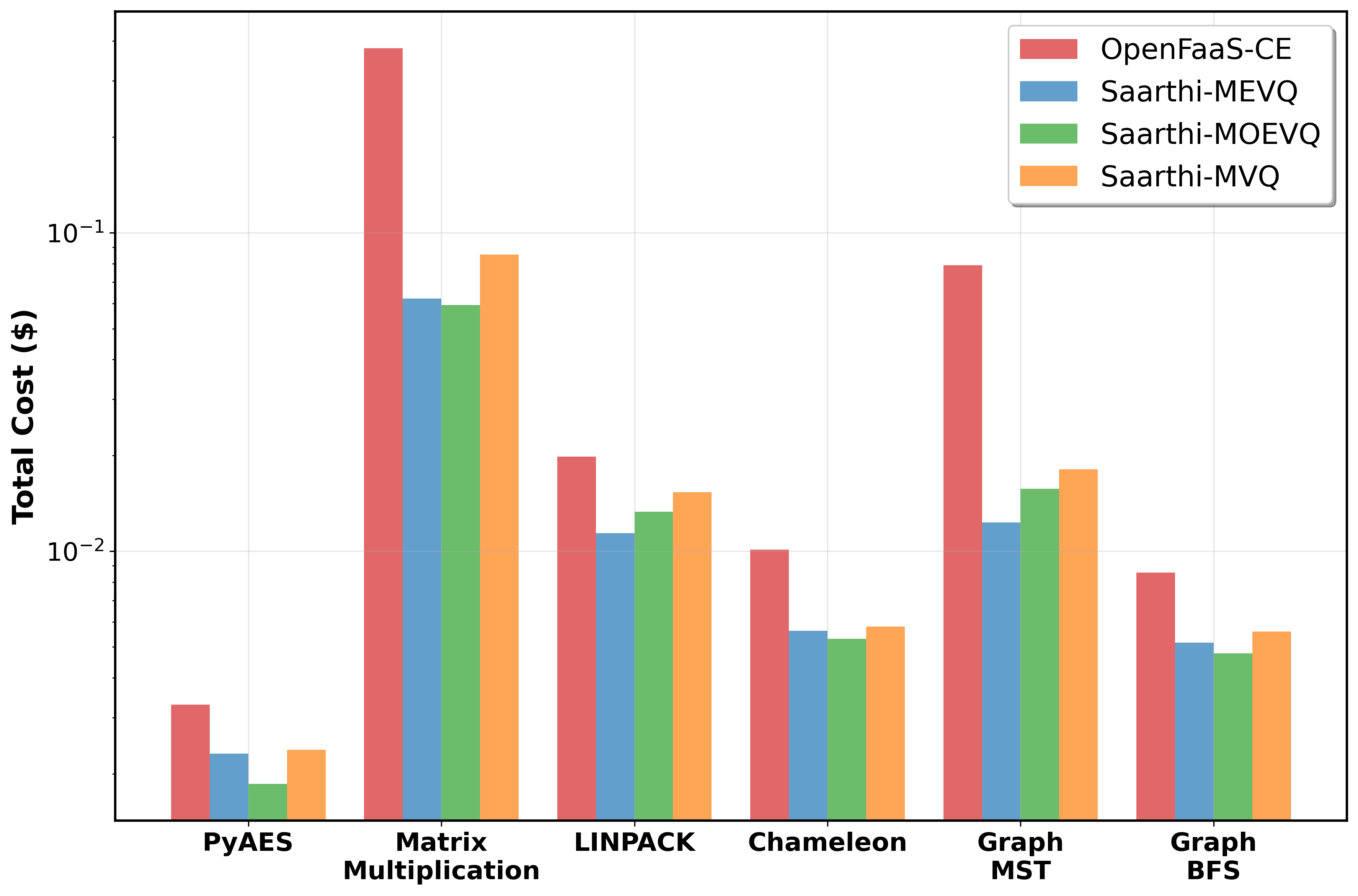}
    \caption{Operational Cost Comparison of Different Variants}
    \label{fig:cost_reduction}
\end{figure}

\begin{figure}[h]
    \centering
    \includegraphics[width=1\linewidth]{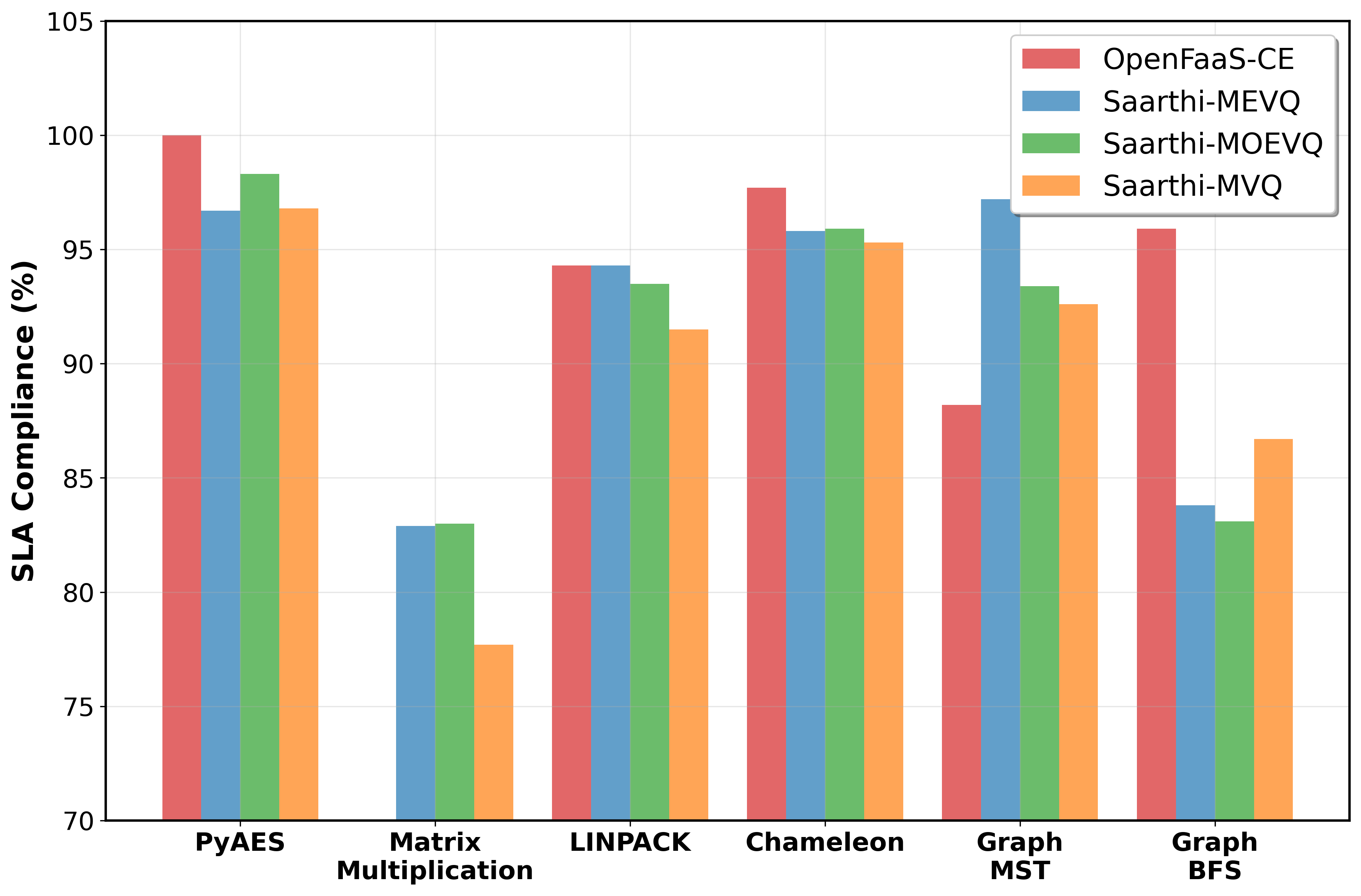}
    \caption{Execution Time SLA comparison of Saarthi variants as compared to OpenFaaS-CE}
    \label{fig:sla}
\end{figure}

\begin{figure}[h]
    \centering
    \includegraphics[width=1\linewidth]{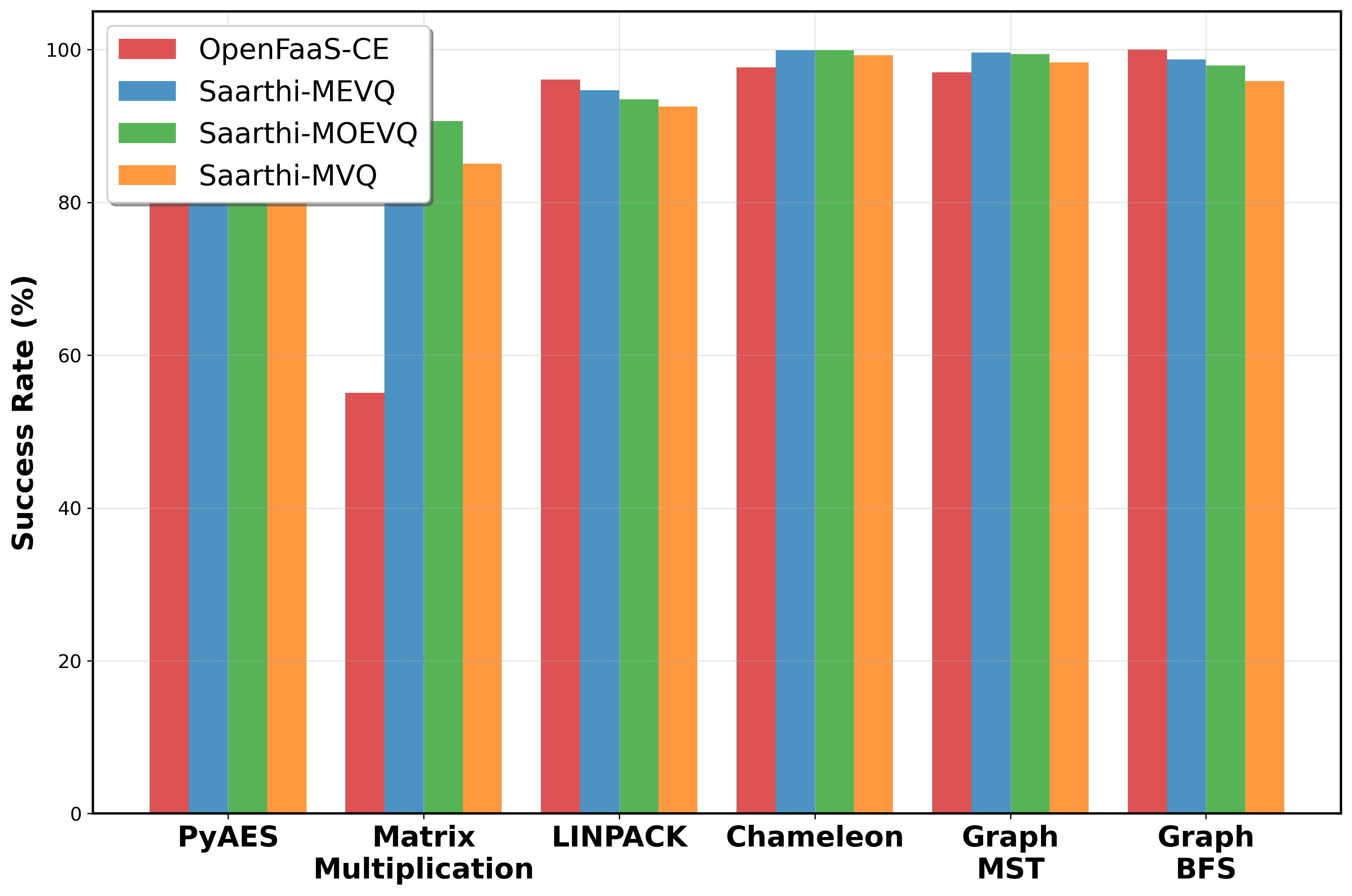}
    \caption{Request Success Rate Analysis per Workload for OpenFaaS-CE and Saarthi variants}
    \label{fig:success}
\end{figure}

\begin{figure}[h]
    \centering
    \includegraphics[width=1\linewidth]{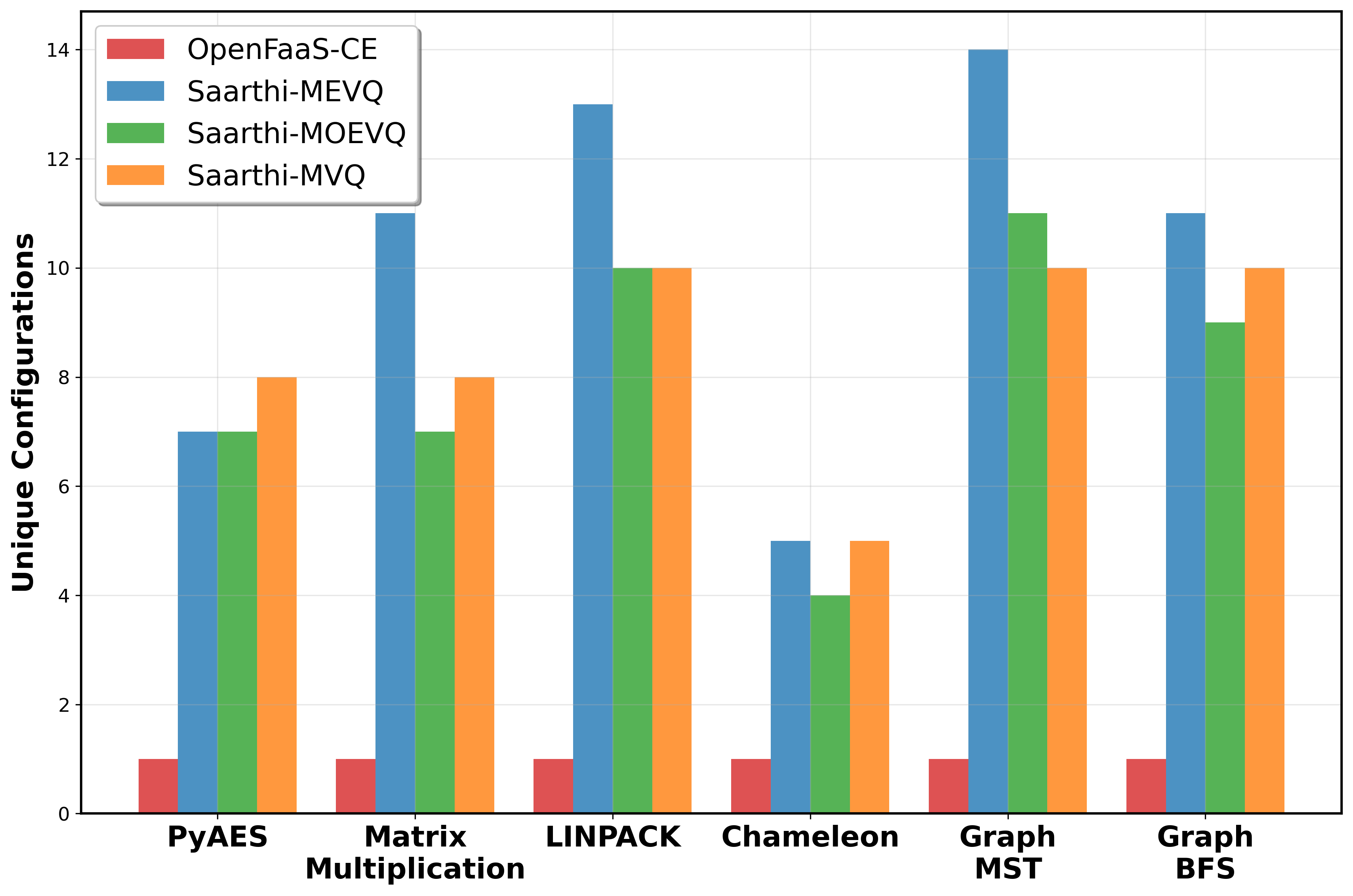}
    \caption{Unique Configurations used by OpenFaaS-CE and Saarthi variants}
    \label{fig:configs}
\end{figure}
\begin{figure}[h]
    \centering
    \includegraphics[width=1\linewidth]{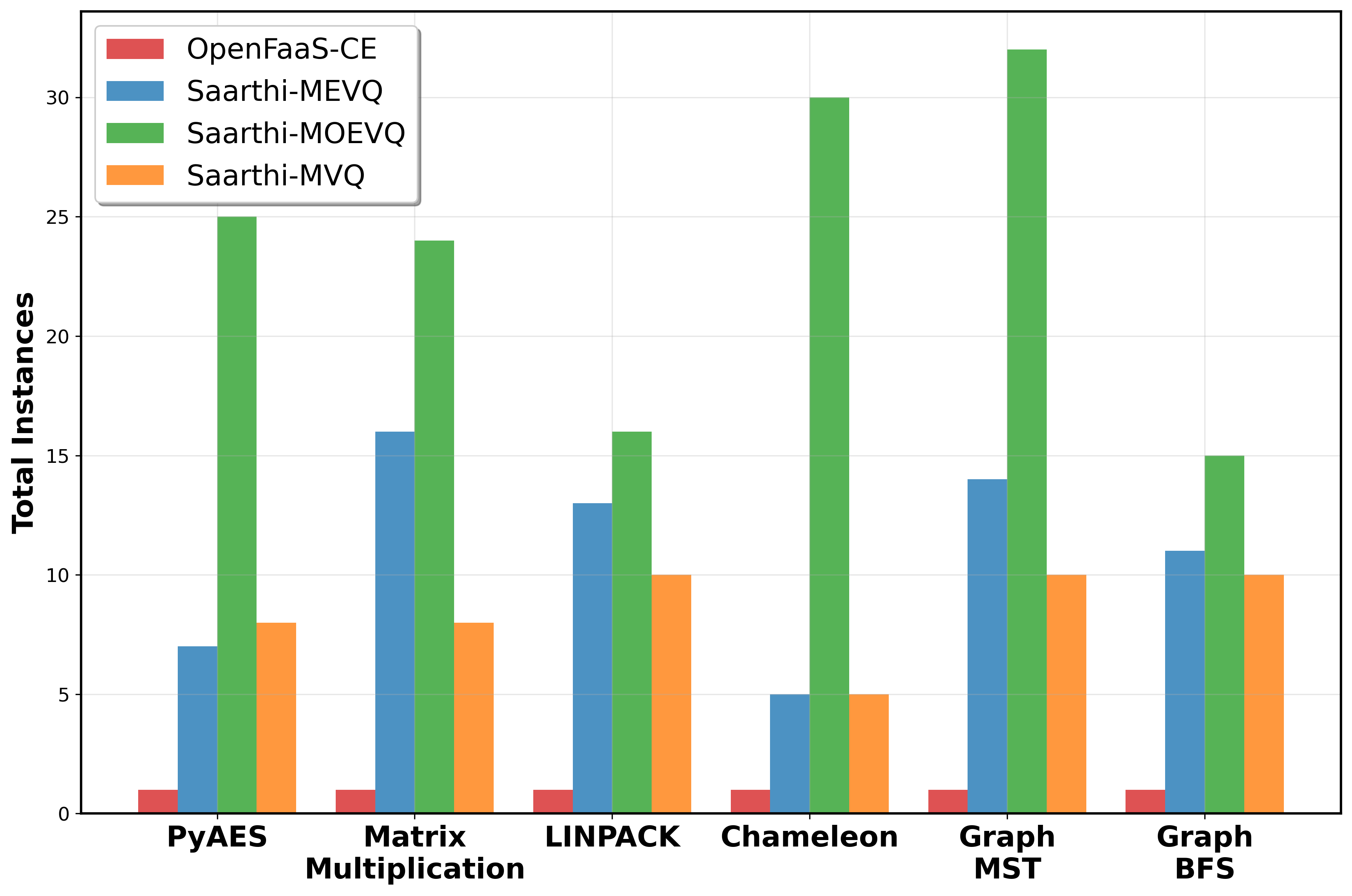}
    \caption{Total Instances used by OpenFaaS-CE and Saarthi variants across experiments}
    \label{fig:instances}
\end{figure}

\paragraph{Results}
\label{para:results}
Our comprehensive experiments across six diverse workloads reveal significant performance improvements and cost optimisations achieved by the Saarthi framework and its variants. We summarise the comparison results in Fig. \ref{fig:cost_reduction}-\ref{fig:instances} and focus on execution latency, operational cost, and resource usage across the entire workload. We observe for \textit{graphmst} with an \textit{orchestration} trigger that Saarthi variants perform better, with Saarthi-MOEVQ and Saarthi-MEVQ achieving up to 93.4\% and 97.2\% execution latency SLA satisfaction compared to 88.2\% for the baseline OpenFaaS-CE (Fig. \ref{fig:sla}). Furthermore, Saarthi-MEVQ successfully serves 0.2\% more workload than Saarthi-MOEVQ while saving up to 21.2\% of operational costs (Fig. \ref{fig:cost_reduction}). This reduction is further magnified for Saarthi-MVQ, which incurs a 1.32x times operational cost while satisfying 92.6\% execution time SLA, and costing 1.84x times for OpenFaaS-CE, as compared to the best-performing Saarthi variant. In terms of resource usage, Saarthi-MOEVQ leverages 11 different function configurations compared to 14 by Saarthi-MEVQ (Fig. \ref{fig:configs}, which introduces more unique instances during the experiment due to its optimisation-based instance management, Fig. \ref{fig:instances}. Similar results are observed for \textit{graphbfs} where Saarthi-MEVQ satisfies 0.7\% more execution SLA as compared to Saarthi-MOEVQ with 7.2\% extra operational cost. Additionally, Saarthi-MOEVQ explores 18\% less function configurations while exploiting 36\% more unique instances. Contrastingly, for \textit{chameleon} with an \textit{http} trigger, OpenFaaS-CE is unable to handle a burst of requests, resulting in 96\% more failures, Fig. \ref{fig:success} and a 1.47x and 1.3x times more expensive execution than Saarthi-MOEVQ and Saarthi-MEVQ, respectively. However, OpenFaaS-CE is able to satisfy 1.18x times more execution time SLA out of successfully served requests due to its over-provisioned resources. OpenFaaS-CE did not scale any instances, but Saarthi-MOEVQ utilised 4 different versions while using 30 unique instances over the experiments.

\begin{figure}[h]
    \centering
    \includegraphics[width=1\linewidth]{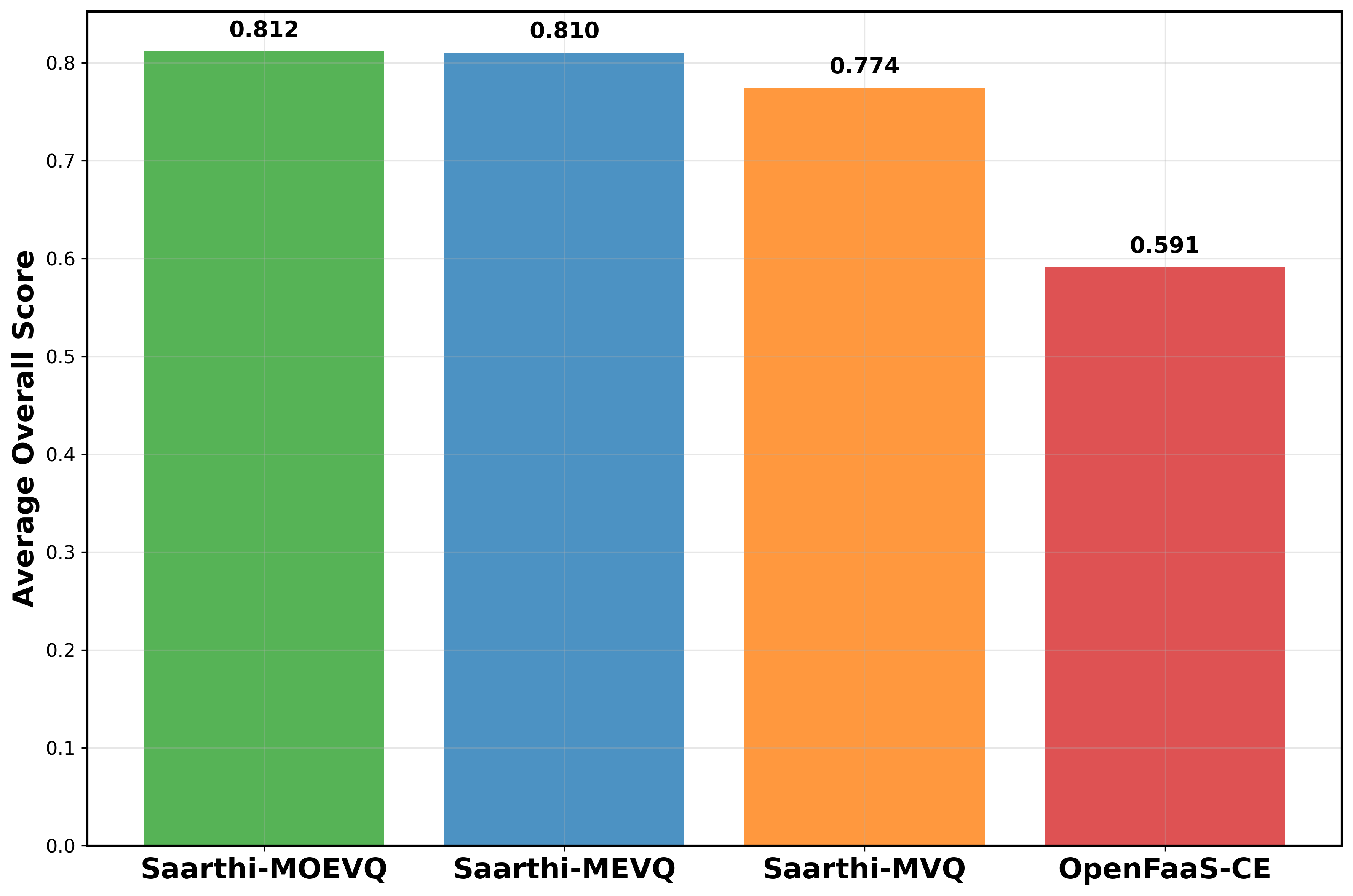}
    \caption{Average Overall Performance Score of OpenFaaS-CE and Saarthi variants}
    \label{fig:performance}
\end{figure}

For resource-intensive functions such as \textit{matmul} and \textit{pyaes}, Saarthi outperforms the baseline with a larger margin. For \textit{matmul}, Saarthi-MOEVQ achieves 83\% execution time SLA satisfaction under a heavy and bursty workload while OpenFaaS-CE struggles to keep up with the incoming requests, serving only around 42\%, Fig. \ref{fig:success}. Additionally, the operational cost surpasses up to 84\% for OpenFaaS-CE as compared to both Saarthi-MOEVQ and Saarthi-MEVQ. This further highlights the advantage of the proposed optimisations where up to 24 unique instances are used across 7 different versions by Saarthi-MOEVQ while scaling 3 of them. This exploration is 36\% less for Saarthi-MOEVQ as compared to Saarthi-MEVQ, where both account for at most 3 function version failures. In the case of \textit{pyaes}, even though Saarthi-MOEVQ satisfies 1.7\% less execution time SLA, the trade-off between cost savings and SLA are higher with savings of up to 43\% as compared to baseline OpenFaaS. Consequently, Saarthi variants are able to explore up to 8 different input-aware function versions, using up to 25 unique instances across the test for Saarthi-MOEVQ. Similar results are achieved for \textit{linpack}, where the baseline and Saarthi variant satisfy up to 94.3\% execution time SLA with a reduction of operational costs of up to 42\%. In this case, Saarthi-MEVQ outperforms Saarthi-MOEVQ by 0.8\% with a cost savings of up to 14.3\%, however, exploring 30\% more function versions and 20\% less unique instances. Therefore, we observe a superior overall performance (normalised weighted sum of SLA, cost and success rate), Fig. \ref{fig:performance}, of Saarthi variants as compared to OpenFaaS-CE, with Saarthi-MOEVQ leading at a score of $0.812$.


\paragraph{Overhead Analysis}
\label{para:overheads}

In addition to the performance boost that Saarthi provides over the baseline, the overheads introduced by different components are minimal and comparable to the short execution of functions. The critical strategies of various components, such as the adaptive request balancer, optimisation engine, and fault-tolerance mechanism, are input-aware predictions, ILP-based cluster optimisation, finding alternate and idle function versions suitable for successful execution, and applying scaling decisions. The prediction service has an overhead of 0.1 seconds for a unique inference that is followed by 0.1 milliseconds for cached inferences. Once the resource requirement is ascertained, the adaptive balancer incurs an overhead of an average of 40 milliseconds to determine available idle function instances or any alternate versions. In doing so, it might explore the possibility of a cold start, and applying that decision takes 0.2 seconds (part of the OpenFaaS-CE overheads).Similarly, the redundancy-based fault-tolerance mechanism also experiences an overhead of 0.2 seconds to apply its decisions. The independent ILP optimisation service that executes every minute to balance the cluster state takes 1.45 seconds to run a single optimisation loop, and the decisions are applied based on the overheads incurred by the base OpenFaaS backend. However, the cold-start overhead introduced by the Kubernetes in our setup varies between 2 to 6 seconds, depending on the cached function image at the underlying nodes.

\subsection{Discussion and Lessons}
\label{sec:discussion}

Based on the results we have observed, we can clearly realise the benefit of Saarthi's components and its input-aware approach. However, the performance variation between Saarthi-MOEVQ and Saarthi-MEVQ highlights a crucial trade-off. The Saarthi-MOEVQ variant, with its ILP optimisation engine, proactively seeks an optimal state, which can sometimes result in cold-starts when a non-existent function version is predicted. As discussed, the cold start overheads may vary from 2 seconds to 6 seconds, this not only leads to a higher rate of failed requests but also contributes to increased execution times, thus reducing the overall SLA compliance. However, these performance drawbacks can be addressed by tuning the platform for specific applications and by focusing on improvements to cold-start times. On the other hand, Saarthi-MEVQ does not employ a scale-down mechanism and continues to operate with a larger number of available function instances, which were added to compensate for earlier failures. This strategy ensures higher reliability and SLA compliance at the expense of operational cost. Conversely, in cases where OpenFaaS-CE is able to serve more traffic and achieve higher SLA rates, this is often attributed to its static configuration and over-provisioned resources. This results in faster and more SLA compliant executions but at a significantly higher and often unnecessary cost. Our comprehensive experiments across diverse workloads reveal significant performance improvements and cost optimisations achieved by the Saarthi framework and its variants. For compute-intensive functions like \textit{matmul} and \textit{pyaes}, Saarthi-MOEVQ already performs above par, demonstrating the clear benefits of its optimisation engine. An improvement in cold-start times would further benefit Saarthi-MOEVQ's performance for other lightweight functions, making its performance advantages more widespread across a variety of workloads.

Our implementation decisions were guided by a series of deliberate trade-offs to ensure the framework's effectiveness within a Kubernetes-based environment. The hybrid orchestrator, for instance, employs a simple, difference-based scoring heuristic to maintain fast decision-making. The exploration probability for cold-starts was set at 20\% to avoid being overly restrictive while still allowing for the discovery of a better long-term state. To prevent resource thrashing, the redundancy-based fault-tolerance approach was configured with a 30-second cooldown period, running checks every 15 seconds to ensure a balance between responsiveness and stability. The ILP optimiser's decisions are constrained by a configurable average function throughput of 10 requests per minute and a pricing scheme based on AWS pricing \cite{awsLambdaPricing}. To match the behaviour of OpenFaaS-CE, the optimiser does not scale down to zero.

The experimental setup was carefully designed to provide a realistic evaluation. The 2-hour test duration was selected to be consistent with the 2-hour model refresh window of our prediction service, MemFigLess. The request queue within the \textit{faas-netes} provider was configured with a capacity of 10 requests and a 10 milliseconds wait time, aligning with the provider's native retry behaviour. We also accounted for the inherent overheads of the underlying platform in ILP optimisation cycles, for example, a cold start takes between 2-6 seconds, depending on whether the image is cached. We found that despite our system's added complexity, the overheads are minimal and acceptable. The latency added by our components typically adds at most 0.2 seconds to the critical path of a cold start. This minimal overhead is a worthwhile trade-off given the substantial cost savings and SLA improvements observed across the board.

\begin{table} 
\centering
\resizebox{\columnwidth}{!}{%
\begin{tabular}{|l|p{3cm}|p{4cm}|}
\hline
\textbf{Paper} & \begin{tabular}[c]{@{}l@{}}\textbf{Focus Area}\end{tabular} & \begin{tabular}[c]{@{}l@{}}\textbf{Key Contribution}\end{tabular} \\ \hline
\cite{UCCSiddharth} & Random forest regression & Input-aware prediction, online pipeline \\ \hline
\cite{ParrotFish} & Online regression & Memory configuration recommendation, not input-aware \\ \hline
\cite{StepConf} & SLO-aware conf. & Balances parallelism for workflow SLO \\ \hline
\cite{Llama}, \cite{charmseeker}, \cite{Astrea} & Video analytics, pipeline tuning & Input-aware configuration, processing time/cost \\ \hline
\cite{OFC} & Opportunistic caching & Input-aware in-memory cache performance \\ \hline
\cite{COSE.V2} & BO & Near-optimal memory configuration, sample-efficient \\ \hline
\cite{MAFF} & Memory search (linear/binary) & Optimises function memory for cost and time \\ \hline
\cite{ChunkFunc} & Input-dependent performance, BO & Profiling pipeline, workflow memory optimisation \\ \hline
\cite{Cypress} & Input-sensitive scheduling & Predictive scaling, batching, resource/energy use \\ \hline
\cite{Pheromone} & Data-centric scheduling & Expresses data patterns, two-tier orchestration \\ \hline

\cite{Orion} & Sizing, bundling, pre-warm & Cold start reduction in Directed Acyclic Graphs (DAG) \\ \hline
\cite{Jiagu} & Migration, placement & Optimises placement, migration under practical constraints \\ \hline
\cite{Fifer} , \cite{Ensure} & Queues, pre-warming, scaling & Reduces cold starts, manages contention \\ \hline
\textbf{Saarthi (Our Work)} & Routing, Scaling and Scheduling & Optimises request orchestration, Function re-use, Optimises Cluster state \\ \hline
\end{tabular}%
}
\caption{Summary of Related Works}
\label{tab:related-work}
\end{table}

\section{Related Work}
\label{sec:related_work}

A number of existing studies have explored the resource configuration aspect in FaaS for both multi- and single function applications, however, focusing on distinct tunable parameters and objectives. The researchers \cite{MAFF} employ various search methods such as linear or binary search to find the right function memory configuration optimised for operational cost and execution time. Another work \cite{StepConf} proposes to balance the inter- and intra- function parallelism while achieving the workflow SLO. Alternatively, studies \cite{Llama}\cite{charmseeker}\cite{Astrea} explore the video processing and analytics pipeline, where they discuss the criticality of function inputs and propose configuration tuning frameworks to find optimal or near optimal configurations to improve the processing time and reducing cost. Taking a different approach, \cite{OFC} attempts to reclaim the extra function memory to maintain a data cache based on the input-aware resource prediction. The authors \cite{COSE.V2} exploit Bayesian Optimisation (BO) to learn the relationship between execution time and memory configurations and reduce sample size. In doing so, they propose a framework to find the near optimal function memory while satisfying customer objectives. Complementary to this, Moghimi et al. \cite{ParrotFish} presents an online parametric regression technique to recommend the right memory configuration for functions, however, not focusing on the function input. Consequently, \cite{UCCSiddharth} explores RFR learning for predicting input-aware memory settings, and provides a profiling and inference pipeline for online function scheduling while lowering operational cost and resource wastage. Similarly, ChunkFunc \cite{ChunkFunc} also exploits the input-dependent function performance to build performance models and put forward a BO-based profiling pipeline that is leveraged to optimise the memory configurations for serverless workflows. However, none of the discussed works offer an end-to-end framework that manages the function request lifecycle, starting from resource prediction to smart orchestration, reducing resource wastage and cold starts to finally optimising the function deployments for determining right balance of instance counts.  

In addition to resource configuration works, a significant effort has been spent by the community on function scheduling and placement \cite{SchedulingSurvey}\cite{scheduling2018}, exploiting distinct aspects such as data locality \cite{Pheromone}, workload interference \cite{UnderstandingInterference}, cold starts \cite{Orion} and migration \cite{Jiagu}. Bhasi et al. \cite{Cypress} introduce input-size sensitive request re-ordering and batching to determine the number of function instances. They propose a set of predictive scaling components that support the request scheduling to improve cluster resource utilisation and energy consumption. Zhao et al. \cite{UnderstandingInterference} proposes a QoS and performance predictor for co-located services and schedule functions under partial interference. Gunasekaran et al. \cite{Fifer} introduce queues for the incoming requests to reduce the number of function instances and pre-warm instances using Long Short Term Memory (LSTM)-based prediction models. Similarly, \cite{Ensure} also addresses the resource contention experienced by a diverse set of functions and employ a scheduling and scaling component to regulate the resource usage of instances and prevent cold starts. Pheromone \cite{Pheromone} follows a data-centric approach where data drives the function execution in a workflow. They introduce new primitives to express data consumption patterns and function orchestration in order to reduce the data transfer and utilise a two-tier coordination to execute functions. Although, these works address different aspects of function scheduling, none of them directly focus on input- and version-aware function scheduling and management. A summary of related works is listed in Table \ref{tab:related-work}.


\section{Conclusions and Future Work}
\label{sec:conclusions}

In this paper, we presented \textbf{Saarthi}, an end-to-end framework designed to address the inherent challenges of resource management in FaaS platforms. Our work explores beyond static, developer-driven configurations and simple reactive scaling, demonstrating a practical approach to building a \textit{self-driving} serverless platform. The core of Saarthi lies in its harmonised components such as an adaptive request balancer that intelligently orchestrates per-request decisions, a proactive ILP-based optimisation engine that manages long-term cluster state, and a resilient fault-tolerant redundancy mechanism that mitigates service failures. Through a comprehensive comparative study, we demonstrated that Saarthi and its variants significantly outperform the baseline OpenFaaS-CE, achieving substantial improvements in key metrics. The hybrid orchestration strategy effectively managed dynamic and bursty workloads, resulting in lower end-to-end latency and higher SLA compliance. Furthermore, the ILP optimiser proved its ability to reduce operational costs and resource fragmentation by making globally optimal decisions. The redundancy mechanism was shown to be effective at ensuring service continuity by preventing sustained failures. 

While Saarthi presents a significant improvement, several promising directions for future work have been identified. First, while the optimisation engine can currently query a prediction service, its default greedy mode remains reactive. Future work could focus on making the predictive model the primary driver of the optimiser, allowing for more proactive and efficient resource management. This could also involve exploring more complex algorithms to dynamically determine the optimal instance configurations and scaling decisions. Second, the adaptive orchestrator currently uses a fixed probabilistic exploration window of 20\%. A significant area for future research is to develop an adaptive exploration model, where this parameter dynamically adjusts based on observed system performance, workload patterns, and resource availability. Finally, while the current redundancy mechanism effectively compensates for common instance failures such as out-of-memory, it does not account for a wider range of failures like node failures, in-flight request failures, etc. Future work could focus on expanding the fault-tolerance mechanism to detect and compensate for more complex scenarios, further strengthening the framework's resilience.

\section*{Acknowledgments}
\label{sec:ack}
This work is supported by the Australian Research Council (ARC) grants: Discovery Project (DP240102088) and  Linkage Infrastructure, Equipment and Facilities (LE200100049).

\bibliographystyle{plain}
\bibliography{references}

\end{document}